\documentclass[sigconf]{acmart}
\usepackage[utf8]{inputenc}

\usepackage{soul}
\usepackage{xcolor}
\usepackage{xspace}
\usepackage{url}
\usepackage{graphicx}
\usepackage{listings,multicol}
\usepackage[caption=false]{subfig}  
\usepackage[export]{adjustbox}
\usepackage{amsmath}
\usepackage{textcomp}
\usepackage{breakurl} 
\usepackage{cleveref}
\usepackage{algorithm}
\usepackage[noend]{algpseudocode}
\usepackage{diagbox}
\usepackage{fixfoot}

\usepackage{balance}

\newcommand{\eg}{\emph{e.g.,}\xspace}

\DeclareFixedFootnote*{\review}{The original links cannot be available for the purposes of the double-blind review. Currently, anonymized versions of the tool and the Docker image can be found in the supplementary materials~\cite{artifacts}. The original links will be made public in the camera-ready version.}

\copyrightyear{2022}
\acmYear{2022}
\setcopyright{acmlicensed}\acmConference[SIGCSE 2022]{Proceedings of the 53rd ACM Technical Symposium on Computer Science Education V. 1}{March 3--5, 2022}{Providence, RI, USA}
\acmBooktitle{Proceedings of the 53rd ACM Technical Symposium on Computer Science Education V. 1 (SIGCSE 2022), March 3--5, 2022, Providence, RI, USA}
\acmPrice{15.00}
\acmDOI{10.1145/3478431.3499294}
\acmISBN{978-1-4503-9070-5/22/03}
\acmDOI{10.1145/3478431.3499294}

\newcommand{\toolname}{\emph{Hyperstyle}\xspace}

\newcommand{\secpart}[1]{\subsection{#1}}

\title{\toolname: A Tool for Assessing the Code Quality\\of Solutions to Programming Assignments}

\author{Anastasiia Birillo}
\affiliation{
  \institution{JetBrains Research}
}
\email{anastasia.birillo@jetbrains.com}

\author{Ilya Vlasov}
\affiliation{
  \institution{Saint Petersburg State University}
}
\email{ilyavlasov2011@gmail.com}

\author{Artyom Burylov}
\affiliation{
  \institution{Stepik}
  \institution{Miro}
}
\email{avburylov@gmail.com}

\author{Vitalii Selishchev}
\affiliation{
  \institution{Computer Science Center}
}
\email{vvselishchev@gmail.com}

\author{Artyom Goncharov}
\affiliation{
  \institution{Computer Science Center}
}
\email{artyom.goncharov1@gmail.com}

\author{Elena Tikhomirova}
\affiliation{
  \institution{JetBrains Research}
}
\email{elena.tikhomirova@jetbrains.com}

\author{Nikolay Vyahhi}
\affiliation{
  \institution{Stepik}
}
\email{vyahhi@stepik.org}

\author{Timofey Bryksin}
\affiliation{
  \institution{JetBrains Research}
  \institution{Saint Petersburg State University}
}
\email{timofey.bryksin@jetbrains.com}

\begin{abstract}
In software engineering, it is not enough to simply write code that only works as intended, even if it is free from vulnerabilities and bugs. 
Every programming language has a style guide and a set of best practices defined by its community, which help practitioners to build solutions that have a clear structure and therefore are easy to read and maintain. 
To introduce assessment of code quality into the educational process, we developed a tool called \toolname. 
To make it reflect the needs of the programming community and at the same time be easily extendable, we built it upon several existing professional linters and code checkers.
\toolname supports four programming languages (Python, Java, Kotlin, and Javascript) and can be used as a standalone tool or integrated into a MOOC platform. We have integrated the tool into two educational platforms, Stepik and JetBrains Academy, and it has been used to process about one million submissions every week since May 2021. 
\end{abstract}



\begin{document}
\fancyhead{}

\maketitle

\section{Introduction}\label{sec:introduction}

Code quality is an important aspect in software development~\cite{martin2009clean}. 
Poor coding style may result in writing incomprehensible code that is difficult to maintain and test~\cite{glass2002facts}.
Creating code quality awareness is an important step in preparing programming students to work as professional developers~\cite{keuning2017code, borstler2018know, keuning2019teachers}.
However, many students do not pay enough attention to code quality, since the main goal for them is to submit a solution that passes all the tests~\cite{keuning2017code}.
They rarely research more complex solutions or best practices~\cite{keuning2017code, keuning2019teachers} and may make the same mistakes over again. 
Therefore, learners may need an incentive to improve their coding style~\cite{keuning2017code}.

Code quality can be checked manually or with special tools.
During manual evaluation, the teacher takes into account the task context and difficulty, and can provide personalized feedback~\cite{keuning2019teachers}.
However, it scales rather poorly. 
Programming tasks in massive open online courses (MOOC) cannot be assessed manually because of the sheer amount of resources needed to adequately evaluate each student's performance~\cite{sakermanoverview}. 
Although there exist a lot of tools for automatic code quality assessment (linters), most of them have not been adapted to the learning process: they aim to detect bugs~\cite{hangal2002tracking}, code smells~\cite{fowler2018refactoring}, or vulnerabilities~\cite{jimenez2009software} in small to large codebases rather than individual projects or single-file solutions.
Such tools usually have high threshold values in their default settings, so minor issues such as small duplicated blocks of code are usually not reported~\cite{keuning2019teachers}. 
Moreover, professional code analysis tools naturally do not aim to track students' progress or provide detailed educational feedback which is crucial for the educational process~\cite{keuning2019teachers}. 

On the other hand, existing research tools~\cite{blau2015frenchpress, choudhury2016scale, ureel2019automated, keuning2021tutoring} designed for assessment of solutions to programming assignments do include only relevant checks and provide feedback which is detailed enough for learners.
However, most such tools support just one programming language and a limited number of embedded tasks, mainly for beginners. 
Finally, several researchers have suggested that the better way of creating code quality assessment tools is not to write one's own validators but to reuse inspections of several real-world analyzers~\cite{rutar2004comparison} and adapt them to the educational process~\cite{keuning2019teachers}.

In this work, we propose \toolname~\cite{hyperstyle}, a tool for automated assessment of the code quality of programming solutions that are written in Python, Java, JavaScript, and Kotlin.
With this tool, we aim to find areas of improvement in programs that pass all the necessary tests but still have weaknesses in terms of readability, maintenance, and complexity.

\toolname adapts existing professional code analyzers to the content and goals of programming assignments. 
In addition, it takes into account students' history of solutions to indicate repeated mistakes.
All code quality issues have a difficulty level assigned. 
Thus, a teacher or a student can get only code quality issues that satisfy the desired level of programming experience.
\toolname is flexible and can be extended by other researchers and practitioners according to their needs. 

We integrated \toolname into two online educational platforms: Stepik~\cite{stepik} and JetBrains Academy~\cite{hyperskill}. 
Within these platforms, more than one million code fragments are submitted by students each week. 
Previously, these platforms only used an automated task validation system that indicated how many tests were passed for a particular submission, but with \toolname the feedback was enriched also with code quality reports.

To sum up, with this work we make the following contributions:
\begin{itemize}
    \item We implement \toolname, a tool for automated evaluation of the code quality of solutions to programming assignments that could be used in MOOCs to provide detailed adapted feedback to programming students. The tool mainly targets Python and Java, but can also work with Kotlin and JavaScript. Its source code is open and available on GitHub~\cite{hyperstyle}. To simplify the deployment process, we also provide a Docker image~\cite{hyperstyleDocker} with the tool. All research artifacts and supplementary materials are available online~\cite{artifacts}.
    \item We provide a dataset of 107 solutions to six programming tasks in Java created by 17 students with different programming experience. 
    We use this dataset to compare the performance of \toolname with \textit{Tutor} ~\cite{keuning2021tutoring}, the only similar tool with an open implementation.
    \item We evaluate the tool's impact on real students
    from the Stepik and JetBrains Academy education platforms by comparing the median number of code quality issues of 300 Python and 294 Java learners before and after the tool's integration. 
    In total, we collected 24,250 submissions for Python and 5,192 for Java. This dataset is also open and available~\cite{artifacts}.
\end{itemize}

\section{Background}\label{sec:background}

In this section, we provide an overview of professional code quality assessment tools and discuss key points related to adapting such tools to education. 

\secpart{Professional Tools}
The first group of code quality assessment tools is used in the development of software projects and provides warnings about typical problems. In some cases, the tools propose relevant code fixes as well.
They can be divided into several categories:
\begin{enumerate}
    \item Code quality checks and automated fixes provided by integrated development environments (IDEs), such as \textit{IntelliJ IDEA}, \textit{Visual Studio}, or \textit{Eclipse}.
    \item Analyzers (linters), \eg \textit{flake8}~\cite{flake8} or \textit{Pylint}~\cite{pylint} for Python,  \textit{PMD}~\cite{pmd} or \textit{Checkstyle}~\cite{checkstyle} for Java. They work fast, but are usually language-specific and do not perform deep checks, for instance, involving control flow or data flow analysis. However, such tools are used pervasively across the industry because they are easy to integrate, trustworthy~\cite{sadowski2018lessons}, and provide actionable output.
    \item Code security and quality platforms, such as \textit{Codacy}~\cite{codacy}, \textit{SonarQube}~\cite{sonarqube}, or \textit{Qodana}~\cite{qodana}. These tools allow software developers to integrate project-level checks locally, on build servers, or other remote resources.
\end{enumerate}

The main goal of professional tools is to prevent the appearance of inefficient, complex, or vulnerable code~\cite{rutar2004comparison}. 
However, most checks performed by such static analysis tools cannot be applied directly to the assessment of solutions to programming assignments.
The main reason is that students' solutions are usually quite small and therefore do not require most checks designed to ensure the validity and security of industry-level projects~\cite{sakermanoverview, akcurastatic, keuning2017code}.
For example, such errors can be related to performance in high-loaded systems and are very difficult to fix in the education process. So it seems that checks such as \texttt{AvoidFileStream}\footnote{\texttt{AvoidFileStream} inspection in PMD: \url{https://pmd.github.io/pmd-6.36.0/pmd_rules_java_performance.html\#avoidfilestream}} for PMD~\cite{pmd} that allows managing garbage collection should be disabled for student code.
In addition, getting a large number of complex code quality issues can overwhelm and discourage students~\cite{keuning2019teachers}.

Another important requirement of the educational setting that is not met by professional code quality assessment tools is tracking progress: students may make the same mistakes over again and disregard coding best practices or programming language features~\cite{keuning2017code}.
Professional tools do not consider such educational aspects because real-world projects usually have other goals.

\secpart{Education-Oriented Tools}

The issue of assessing code quality in education has been actively studied before~\cite{keuning2017code, borstler2018know, keuning2019teachers}.
Several tools have been created that partially reimplement code quality checks from professional solutions that are relevant to learning needs~\cite{blau2015frenchpress, choudhury2016scale, ureel2019automated, keuning2021tutoring}.

\textit{FrenchPress}~\cite{blau2015frenchpress} is an Eclipse plugin that displays adapted messages for a small set of common code style inconsistencies.
\linebreak \textit{WebTA}~\cite{ureel2019automated} is a web-based tool that reports failed tests, errors common to novices, and stylistic issues. 
The tool provides students with adapted feedback that is more appropriate for their level of understanding.

Choudhury et al.~\cite{choudhury2016scale} and, several years later, Keuning et al.~\cite{keuning2021tutoring} proposed two very similar tools: \textit{AutoStyle} and \textit{Tutor}, respectively. 
These tutoring systems let students practice with improving small programs that are already functionally correct. 
The systems are implemented as special environments that display code quality hints step by step and let students improve the code.

To sum up the benefits, assessment tools for education apply checks that are relevant to the students' skill levels and adapt the requirements accordingly.
In addition, all of them adapt the reports to the needs of the learners by providing tips and explanations. 
Some of them allow tracking students' progress.

However, existing education-oriented tools also have common limitations: they are focused on only one programming language each (mostly Java) and often handle only basic errors typical of novice programmers.
Another important drawback is that these tools are difficult to extend for handling more complex errors, since they implement their own validators and do not reuse selected checks from professional analyzers.

\section{Workflow of \toolname}\label{sec:design}

\toolname is implemented as a Python tool. 
It currently targets solutions in Python and Java, but can also work with Kotlin and JavaScript.

\begin{figure}[t]
\centering
    \includegraphics[width=\columnwidth]{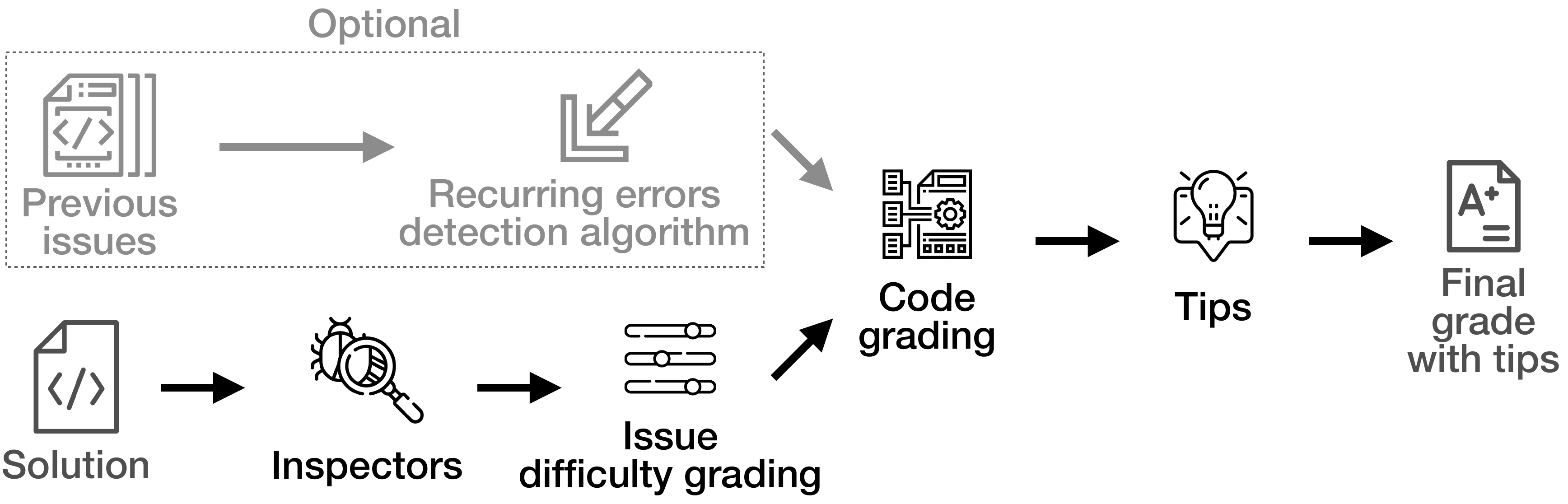}
    \centering
    \caption{Workflow of \toolname.}
    \label{fig:hyperstyle:design}
\end{figure}

Figure~\ref{fig:hyperstyle:design} presents the workflow of \toolname.
The first step of handling a student solution is finding all code quality problems.
Each supported language can have several inspectors.
Then, depending on the selected difficulty level, output only by relevant inspections is taken into account.
After that, all found errors are aggregated and the final grade is calculated.
If a history of the student's previous errors is provided, the current solution can be penalized if some errors are repeated over a certain period of time. 
This step is optional and is disabled if no solution history is provided.
In the end, the least understandable explanations from the analyzers are replaced with more detailed ones.

As input, the tool takes code fragments and optional additional information, \eg programming language, history of previous inspections, and so on. 
\toolname returns the results in the JSON format, which could then be displayed by MOOC platforms in any way suitable: integrated with other user data, highlighted, or omitted if necessary.
The report contains the following information by default: an overall score, a code quality summary, a list of errors by categories with feedback messages, inspections that contributed to penalizing the score (if this happened).
For each error, the tool reports the exact position (number of code line) in the solution.

The rest of this section describes all key concepts and steps of the workflow in more detail.

\secpart{Inspectors and Error Categories}

Code quality issues can be divided into several categories.
In our work, we have identified five main semantic categories of errors.
The categorization is based on the error types used in most of the code inspectors we reviewed and on our own experience:
\begin{itemize}
    \item \textit{Code style}: The code fragment violates one of the rules of the commonly accepted style guide for the chosen language.
    By fixing such issues, students learn to follow popular coding conventions, like PEP-8~\cite{pep8} for Python or Oracle Java code conventions~\cite{oracle} for Java.
    \item \textit{Code complexity}: The solution is poorly designed or overly complicated. By fixing such issues, students learn to make code easier for understanding, editing, and debugging.
    \item \textit{Error-proneness}: The code contains a potential bug. Even if the code passes all automatic tests, it may behave incorrectly in some cases, which would be a problem in a real-world environment. By fixing such issues, students learn to write reliable code.
    \item \textit{Best practices}: The code does not follow the widely accepted recommendations and idioms of the chosen language. Some features of the language can be used in an inefficient or obsolete way. By fixing such issues, students learn to use language features correctly.
    \item \textit{Minor issues}: The code contains problems usually related to incorrect spelling. These problems are worth solving because they hinder the readability of the code. For such errors, the final grade is not reduced.
\end{itemize}
We provide examples of all such error types in the supplementary materials~\cite{artifacts}.

For each language, we selected analyzers that find problems from the categories described above. 
From each analyzer, we manually selected checks that are relevant to typical programming tasks in MOOCs and categorized them.
The list of supported languages, used linters, and the number of unique checks in each error category are provided in Table~\ref{tab:inspectors:stat}.

\begin{table}[htb]
    \begin{tabular}{l|ccccc|c}
    \toprule
    \backslashbox{Language}{Category} & 
    \multicolumn{1}{c}{\begin{tabular}[c]{@{}c@{}}CS\end{tabular}} & \multicolumn{1}{c}{\begin{tabular}[c]{@{}c@{}}CC\end{tabular}} & \multicolumn{1}{c}{\begin{tabular}[c]{@{}c@{}}EP\end{tabular}} & \multicolumn{1}{c}{\begin{tabular}[c]{@{}c@{}}BP\end{tabular}} & \multicolumn{1}{c|}{MI} & \multicolumn{1}{c}{\begin{tabular}[c]{@{}c@{}}Total\end{tabular}} \\ \midrule
    & & & & & & \\ [-0.8em]
    \begin{tabular}[c]{@{}l@{}}\textbf{Python} (flake8, pylint,\\radon)\end{tabular} 
    & 146  & 35 & 162 & 254 & 3 & \textbf{600} \\[0.6em]
    \begin{tabular}[c]{@{}l@{}}\textbf{Java} (Checkstyle, PMD)\end{tabular}         
    & 50  & 8 & 51 & 110 & 3 & \textbf{222} \\[0.5em]
    \textbf{Kotlin} (Detekt)                                                         
    & 70 & 12 & 21 & 75 & 0 & \textbf{178} \\[0.5em]
    \begin{tabular}[c]{@{}l@{}}\textbf{JavaScript} (ESlint)\end{tabular}            
    & 17 & 1 & 15 & 34 & 0 & \textbf{67} \\
     \bottomrule
    \end{tabular}
    \vspace*{1.2mm} 
    \caption{Number of unique error checks per language and category: CS --- Code style, CC --- Code complexity, EP --- Error-proneness, BP --- Best practices, MI --- Minor issues.}
    \label{tab:inspectors:stat}
    \vspace{-7mm}
\end{table}

\toolname is easily extendable. 
Adding support for a new language requires just implementing an inspector module for this language. 
The implementation of each inspector includes code running the linter, parsing its result, and categorizing all errors into the five categories mentioned above.

\secpart{Code Grading}

Four grades are awarded by the tool:
\begin{itemize}
    \item \textit{EXCELLENT} means that the code strictly follows the style guide, does not have complexity or error-proneness issues, and is easy to read and modify.
    \item \textit{GOOD} means that the code is readable and relatively easy to edit, maintain and extend, but still contains some minor code quality issues.
    \item \textit{MODERATE} means that the code follows the style guide only partially, most of language features are not used correctly.  Sometimes the code may be challenging to understand.
    \item \textit{BAD} means that the code is hard to read and modify and probably does not follow the style guide. Also, the solution might have high complexity and might be error-prone.
\end{itemize}
Note that the \textit{EXCELLENT} grade does not guarantee the complete absence of errors.
However, it does certify that the code does not include any \textit{common} code quality errors in this language.

All errors detected by the tool can be divided into two types by how they contribute to the intermediate score within their semantic category:
\begin{itemize}
    \item \textit{Countable}: All instances of such error type are counted. If the number of errors reaches certain threshold values, the respective score is assigned. An example of this kind of errors is the number of places in the code where the \texttt{for} loop can be replaced with \texttt{forEach}.
    \item \textit{Measurable}: The severity of such errors is evaluated within a scale or an interval. Reaching certain threshold values within the scale impacts the score. An example of this kind of errors is the length of lines in the code.
\end{itemize}

The thresholds of values were selected iteratively.
First, we divided each error category into several subcategories and came up with thresholds for them. 
Each subcategory contains only one kind of issues: either only countable issues or only a measurable issue.
Each measurable issue makes its own subcategory and measures the same metric value, \eg metrics \textit{Length of a Code Line} and \textit{Number of Code Lines in a Function} are semantically different.
At the same time, several countable issues can be placed into one subcategory, \eg \textit{\texttt{for} Loop can be Replaced with \texttt{forEach}} and \textit{Unnecessary Local Variable Before \texttt{return}} are both from the one category (\textit{Best practices}). 
All subcategories with their description can be found in the supplementary materials~\cite{artifacts}.
The final grade is calculated as the minimum of the grades for each subcategory.

The initial thresholds were proposed for each subcategory and each language by three experts with more than four years of programming experience each. 
Then the experts ran \toolname on 50 thousand student solutions with the proposed thresholds and randomly checked 100 code fragments to verify that the assessment worked as expected.
After that, several thresholds, mostly from the \textit{Code complexity} and \textit{Error-proneness} categories, were modified.
These thresholds were lowered since students' solutions are smaller and easier than real-world projects.
After that, we integrated \toolname into the Stepik and JetBrains Academy platforms. At this stage, a test group of students reported unreasonably low grades, and the initial group of experts checked each report manually and decided to decrease (or keep) thresholds values.

The final thresholds were evaluated by an empirical analysis of 250 thousand solutions for each language.
In total, one million solutions were analyzed for all four languages.
For each solution, we calculated the metrics for measurable errors and the frequency of countable errors within each subcategory. 
We plotted distributions of these errors by subcategories, indicated the thresholds and checked how many solutions received each of the grades.
If a lot of submissions had a number of issues that was lower or higher than the selected thresholds, the initial group of experts manually looked at several fragments to check whether the boundaries were set too high or too low.
Finally, the threshold values were modified according to this empirical analysis.
Examples of error distribution charts are provided in the supplementary materials~\cite{artifacts}.

To illustrate this process, let us consider several insights about Python solutions. 
In addition to constructing plots, we calculated the percentage of the number of students solutions that correspond to the thresholds and the grades.
For example, a software metric called \textit{Maintainability Index}~\cite{maintainabilityindex} did not influence the grade in most cases (99.29\%) because it was not applicable to students' solutions, which were mostly small in size. 
We also discovered that the distributions of errors from the \textit{Best practices} and \textit{Error-proneness} categories were very similar and probably should have the same thresholds, \eg for the \textit{Best practices} category these values are: EXCELLENT ---   88.3\%, GOOD -- 8.2\%, MODERATE --- 3.4\%, and zero for the BAD grade.
Also, grades for the \textit{Length of Bool Expressions} subcategory had very low thresholds and 97.93\% of solutions had the highest grade in this subcategory.
\secpart{Detection of Recurring Errors}\label{sec:penalty-algorithm}

When solving programming problems, students often make the same mistakes over and over again~\cite{keuning2017code}. We also noticed this while analyzing students' solutions gathered from the Stepik and JetBrains Academy platforms, so we developed an algorithm to detect such recurring errors and show them to the students.
The algorithm analyzes recent solutions in a programming language by a particular student and finds errors that are identical to the ones reported for the current solution.

However, if a student's score is repeatedly decreased, they may become demotivated and upset, since the real reason for repeating an error may be the lack of understanding of this error~\cite{law2010learning}.
To avoid this problem, based on existing analysis and research~\cite{keuning2017code, borstler2018know, SemanticStyle, altadmri201537} we designed different penalty rates for different subcategories of errors.
The final coefficient for each error subcategory is calculated by three criteria:
\begin{itemize}
    \item \textit{Prevalence}: How common the mistake is among students. The most frequent mistakes~\cite{SemanticStyle, altadmri201537} are about code formatting (\eg incorrect brackets or indentation) and best practices of more experienced programmers (\eg using \texttt{enumerate} instead of \texttt{range} in Python).
    \item \textit{Difficulty}: How hard it is to fix the error. Complexity issues turn out to be the most difficult for students~\cite{keuning2017code}. 
    Also, fixing code according to industrial software metrics is often too hard for students~\cite{keuning2017code} since these metrics usually are too abstract or too complex (\eg \textit{Lack of Cohesion of Methods}~\cite{cohesion} in a class).
    \item \textit{Importance}: How important it is to fix the error. 
    Not all issues should be fixed immediately~\cite{borstler2018know, peitek2021program}. For example, such software metric as \textit{Number of Children of a Class} can be mostly ignored in student solutions.
\end{itemize}

The coefficient for each subcategory is calculated as the sum of all three criteria divided by the maximum of them. 
The exact values of these coefficients can be found in our supplementary materials~\cite{artifacts}.
To calculate the final penalty coefficient, we count the number of issues for each subcategory, multiply it by the coefficient for this subcategory, sum these products for all subcategories and normalize the sum.
As a result, the final penalty coefficient ranges from 0 to 1.

The final code quality grade for each subcategory is reduced by one, two, or three levels. 
If this coefficient is in the range of $[0; 0.5)$ then the final grade is not reduced. 
Next, for every $0.2$ points, the grade is decreased by one level.
The grade reduction factors are presented in Table~\ref{tab:penalty:coefficient}.
These thresholds were selected by three experts with programming experience of more than four years. 
Two of them also have teaching experience of more than three years.
The main goal of the thresholds is not to decrease the grade if the number of recurring errors is low or these errors are too difficult for students.

Let us consider an example. Let the initial grade (without a penalty) be \textit{GOOD} and let the penalty coefficient be 0.6. 
According to Table~\ref{tab:penalty:coefficient}, the final grade for the student will be decreased by one level and become \textit{MODERATE}, since $0.6 \in [0.5; 0.7)$.

 \begin{table}[htbp]
     \centering
     \begin{tabular}{l|cccc}
         \toprule
         \backslashbox{Initial\\grade}{Final\\grade} & \rotatebox[origin=c]{90}{EXCELLENT} & \rotatebox[origin=c]{90}{GOOD} & \rotatebox[origin=c]{90}{MODERATE} & \rotatebox[origin=c]{90}{BAD}
             \\ \midrule \\ [-0.8em]
         EXCELLENT & $[0, 0.5)$ & $[0.5, 0.7)$ & $[0.7, 0.9)$ & $[0.9, 1]$ \\
         \\ [-0.8em]
         GOOD &  --- & $[0, 0.5)$ & $[0.5, 0.7)$ & $[0.7, 1]$ \\
         \\ [-0.8em]
         MODERATE & --- & --- & $[0, 0.5)$ & $[0.5, 1]$ \\
         \\ [-0.8em]
         BAD & --- & --- & --- & $[0, 1]$ \\ \bottomrule
     \end{tabular}
    
     \vspace*{1.2mm} 
     \caption{Influence of the penalty coefficient on the final grade.}
     \label{tab:penalty:coefficient}
     \vspace{-7mm}
 \end{table}

\secpart{Difficulty Levels of Errors}\label{sec:issue_difficulty}

We introduced several difficulty levels of code quality issues. 
This way, learners can choose the level appropriate for their current skill set and thus will not be demotivated by requirements that do not match their level. 

We use three levels that correlate to how hard it is to fix a particular code quality issue based on the criteria of prevalence, difficulty, and importance described in Section~\ref{sec:penalty-algorithm}: 
\begin{itemize}
    \item \textit{Easy}. This group mainly consists of formatting issues from the \textit{Code style} category. These problems are prevalent and can be fixed easily~\cite{SemanticStyle, altadmri201537}.
    \item \textit{Medium}. Issues from this group are related to \textit{Best practices} applied by more experienced programmers, \eg using more efficient built-in functions of the programming language.
    \item \textit{Hard}. This level contains \textit{Error-proneness} and \textit{Code complexity} issues. Fixing them requires a certain level of experience and knowledge. 
\end{itemize}

As future work, we plan to come up with an algorithm for determining the difficulty level of a task automatically. 

\secpart{Providing Feedback}

\toolname provides descriptions for all found errors. 
Default output messages of professional code analyzers are often too brief, whereas students may need more thorough explanations~\cite{keuning2019teachers}. 
We came up with custom messages for some of the issues, mainly for the most difficult categories, \eg \textit{Code complexity} or \textit{Error-proneness}.
On the other hand, we preserve standard messages for easy-to-fix issues, which mostly belong to \textit{Code style} and \textit{Best practices} categories.

\section{Evaluation}\label{sec:evaluation}

We evaluated the tool in two different scenarios.
Firstly, we compared our tool with previous work on a dataset of selected student solutions. 
Even though there are several papers introducing similar tools, only one, \textit{Tutor}~\cite{keuning2021tutoring}, was openly available for comparison.
Next, we evaluated the usefulness of our tool in a setting of a programming course within several MOOC platforms: Stepik~\cite{stepik} and JetBrains Academy~\cite{hyperskill}. For that, we compared the median number of code quality issues per student before the tool was integrated into the platforms and after that. 

\secpart{Comparison with Similar Tools}\label{sec:comparison}

\begin{table}
    \begin{tabular}{l|cc}
    \toprule
    \textbf{Code quality issue} & \textbf{\textit{Tutor}} & \multicolumn{1}{c}{\textbf{\toolname}} \\ \midrule
    \texttt{For} loop can be \texttt{forEach}  & 8  & 9  \\ [0.2em]
    Simplify boolean expression & 1  & 1 \\ [0.2em]
    Simplify boolean return    & 1  & 1  \\ [0.2em] 
    \midrule
    Short algebraic operation        & 4  & -  \\ [0.2em]
    Switch \texttt{If Else} branches           & \multicolumn{1}{c}{1} & - \\ [0.2em]
    Replace \texttt{For} to \texttt{While}           & \multicolumn{1}{c}{1} & - \\ [0.2em]
    \midrule
    Parameter assignment check      & \multicolumn{1}{c}{-} & 2 \\\midrule
    \textbf{Total}   & \textbf{16}  & \textbf{13}\\ \bottomrule
    \end{tabular}
        \vspace*{1.2mm} 
        \caption{Number of code quality issues detected by \textit{Tutor} and \toolname. Formatting issues detected by \toolname are omitted since \textit{Tutor} is not able to find them.}
        \label{tab:comparison:issues}
        \vspace{-7mm}
\end{table}

Several similar tools have been introduced in prior work~\cite{blau2015frenchpress, choudhury2016scale, ureel2019automated, keuning2021tutoring}. 
However, the only tool that is publicly available to run is a demo version~\cite{tutor} for \textit{Tutor}~\cite{keuning2021tutoring}, which we used for the comparison with \toolname. 
No other versions of \textit{Tutor}, open or proprietary, are available except for this demo.
In it, \textit{Tutor} can assess code quality only for six built-in tasks written in Java. 
These tasks mostly target beginners and require to implement the body of a given function. 
Since we cannot extend \textit{Tutor}, we had to perform the comparison on these six tasks.
We asked 17 people to solve all of them. 
The programming background of the participants ranged from no experience to several years of using Java in industry. 
We included experienced programmers in this study, because sometimes they also attend computer science MOOCs~\cite{miller2014functional} and code quality assessment tools should suit all types of students.  
In total, for the six tasks, we collected 107 snippets of code. 
We make this dataset publicly available in our supplementing materials~\cite{artifacts}.

We ran \toolname and \textit{Tutor} on each code fragment and compared their output.
From our dataset, \textit{Tutor} failed to grade 56 fragments (about 52\% of the solutions), raising 20 unique types of errors. Almost all of the errors were about unsupported language elements, such as conditional operators or initialization of arrays. 
\toolname managed to grade all the provided code fragments.

Table~\ref{tab:comparison:issues} provides the comparison of the code quality issues found by both tools in the remaining part of the dataset: the 51 fragments (48\%) that \textit{Tutor} processed successfully. 
Both tools found the same three unique issues (ten cases in total by \textit{Tutor} and 11 in total by \toolname).
In addition, \textit{Tutor} found three unique issues (six cases in total) that were not found by \toolname.
\toolname managed to find one additional unique issue (two cases in total) that was not found by \textit{Tutor}.
In total, \textit{Tutor} found 16 code quality issues and \toolname found 13.
On the remaining 56 solutions that \textit{Tutor} was unable to process, \toolname found five more code quality issues (five cases in total). 
On top of that, on the whole dataset \toolname found 54 fragments (about 50\% of the dataset) with 201 formatting issues in total, which appears to be by far the most prevalent type of mistakes found in this comparison. 

\secpart{Influence on Students' Code Quality}

To measure the usefulness of \toolname for students, we compared the dynamics of code quality issues before integrating this tool in the education platforms and after that. Checks by \toolname have been run on every submission, so students saw the reports and became aware of their errors even if they did not take action. 

We gathered datasets with Python and Java solutions from the Stepik and JetBrains Academy platforms, selecting submissions by students who had at least ten solutions either in the period before \toolname was introduced, or after.
We selected only such submissions that the students agreed to make publicly available and anonymized all the data.
For Python, we gathered submissions of 300 students. 
The first part of the dataset collected before introducing \toolname contains 9,843 submissions from 150 students, the second one that was collected after introducing the tool has 14,407 submissions from 150 students. 
For Java, we gathered submissions from 294 students. There are significantly fewer publicly available solutions for Java than Python on the Stepik and JetBrains Academy platforms, so we gathered 2,000 submissions from 145 students before \toolname was integrated and 3,192 submissions from 149 students after that. 
The dataset is available in our supplementary materials~\cite{artifacts}.

\begin{figure}[t]
\centering
    \includegraphics[width=0.8\columnwidth]{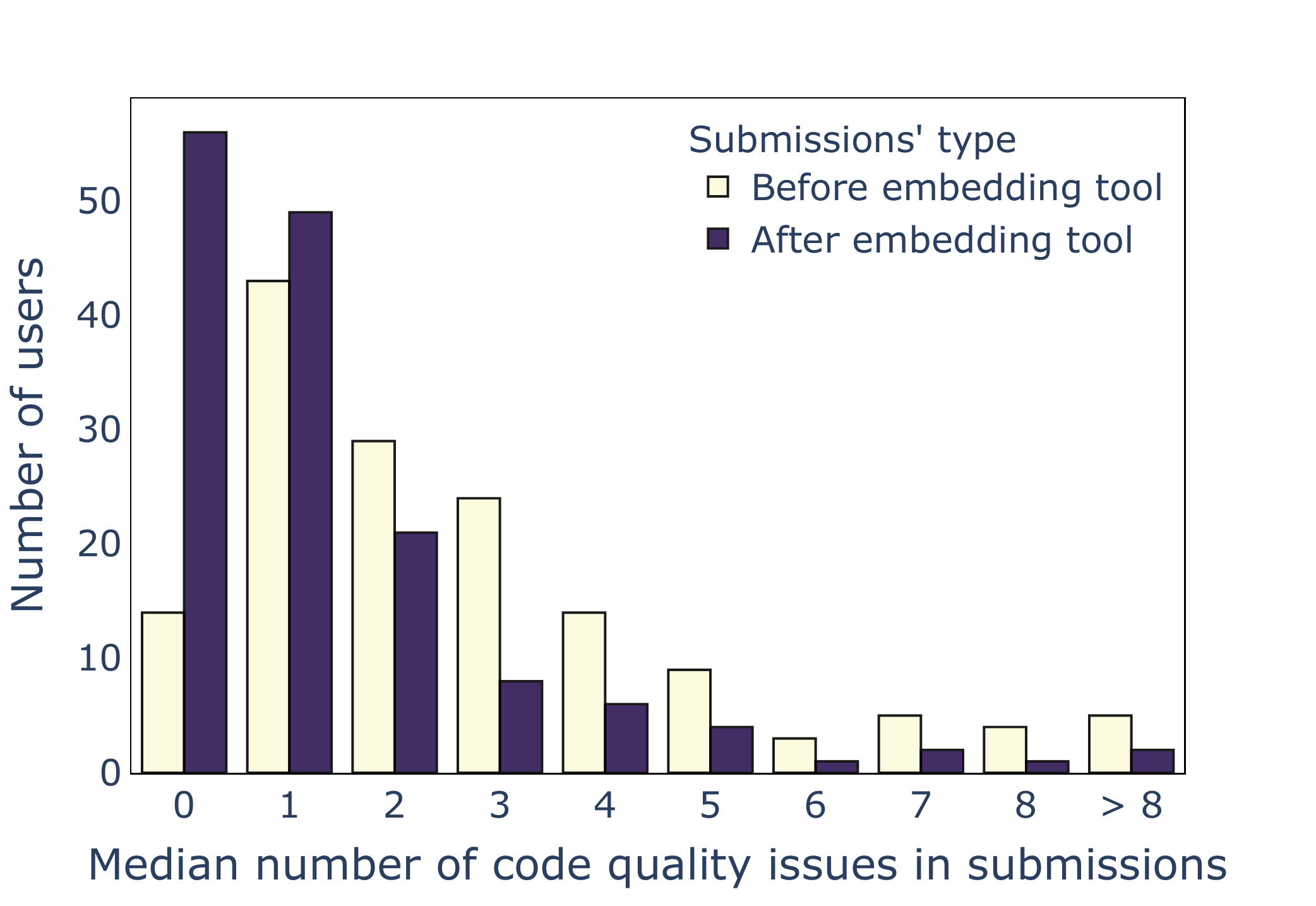}
    \centering
    \caption{Influence of \toolname on students' code style for Python solutions}
    \label{fig:hyperstyle:dynamics:python}
\end{figure}

\begin{figure}[t]
\centering
    \includegraphics[width=0.8\columnwidth]{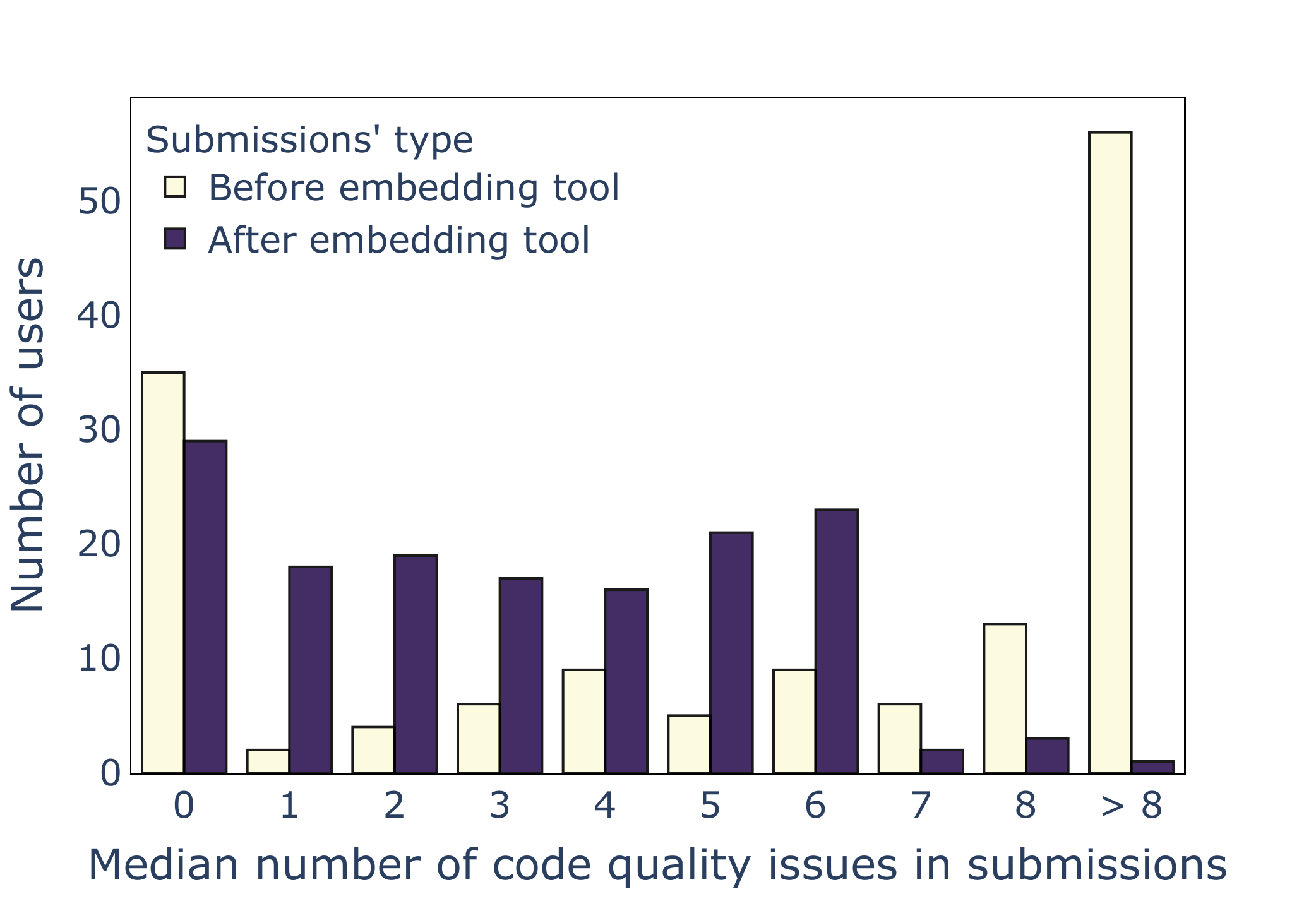}
    \centering
    \caption{Influence of \toolname on students' code style for Java solutions}
    \label{fig:hyperstyle:dynamics:java}
\end{figure}

We ran the tool on all fragments for each student and counted the number of the reported code quality issues.
For each student, we calculated the median number of the issues in their submissions.

Figure~\ref{fig:hyperstyle:dynamics:python} shows how many Python students had a median of 0, 1, etc. code quality issues before and after \toolname was introduced.
These results can be interpreted as circumstantial evidence that the tool contributed to improving the students' code quality on these platforms: the number of students who had no code quality issues increased four times, and the number of students who made two or more errors decreased.

Figure~\ref{fig:hyperstyle:dynamics:java} presents results for Java solutions.
The number of students who made fewer mistakes (one to six) increased, and the number of those who made more than six mistakes, decreased.
The results are especially notable for those who made more than eight errors: their number dropped dramatically from 56 to just one.

\section{Conclusion and future work}\label{sec:conclusion}

In this paper, we introduce \textit{\toolname}---a tool that provides detailed feedback on the code quality of programming solutions, which output can be easily integrated into MOOC platforms. 
The tool currently targets Python and Java, but can also work with Kotlin and JavaScript. 
The tool detects 600 code quality issues for Python and 222 issues for Java divided into five categories: \textit{Code style}, \textit{Code complexity}, \textit{Error-proneness}, \textit{Best practices}, and \textit{Minor issues}.
The tool can also detect recurring issues and thus help students to find issues that repeat over time.
The first version of the tool was launched on the Stepik and JetBrains Academy platforms in January 2020. The current version of \toolname has been running there since May 2021, handling roughly one million submissions every week.

We got very positive feedback from the creators of the Stepik and JetBrains Academy platforms about the quality of the hints, the tool's performance and flexibility.
Since the tool adapted professional linters, students can switch from the online code editors to IDEs more easily since the sets of inspections are very similar.
The most difficult tips were changed into more detailed ones that help students to work with hard code quality issues.

Future work on \toolname involves categorizing code quality issues not only by their semantics and difficulty levels but also priority. 
Apart from that, we are planning to focus on Kotlin and JavaScript languages, since the tool currently supports mostly Python and Java. 
Finally, we are going to develop an algorithm to classify tasks by difficulty so that inspections irrelevant to a task can be disabled automatically.

\bibliographystyle{ACM-Reference-Format}
\balance
\bibliography{cite}

\end{document}